\documentclass[oneside,11pt,a4paper]{report}
\usepackage[english, titlepage]{ku-frontpage}
\usepackage[utf8]{inputenc}
\usepackage{hyperref}
\usepackage{mathtools}

\usepackage{pdfpages}

\usepackage[numbers,sort]{natbib}

\assignment{PhD Thesis}
\author{Casper Hansen}

\title{Representation Learning for Efficient and Effective Similarity Search and Recommendation}
\subtitle{} 
\date{Handed in: April 28, 2021 \\ This thesis has been submitted to the PhD School of The Faculty of Science, University of Copenhagen}
\advisor{Advisors: Stephen Alstrup, Christina Lioma, Jakob Grue Simonsen}

\begin{document}
\maketitle
\pagenumbering{roman}

\phantomsection
\addcontentsline{toc}{chapter}{Abstract}\section*{Abstract}
How data is represented and operationalized is critical for building computational solutions that are both effective and efficient. A common approach is to represent data objects as binary vectors, denoted \textit{hash codes}, which require little storage and enable efficient similarity search through direct indexing into a hash table or through similarity computations in an appropriate space.
Due to the limited expressibility of hash codes, compared to real-valued representations, a core open challenge is how to generate hash codes that well capture semantic content or latent properties using a small number of bits, while ensuring that the hash codes are distributed in a way that does not reduce their search efficiency. 
State of the art methods use representation learning for generating such hash codes, focusing on neural autoencoder architectures where semantics are encoded into the hash codes by learning to reconstruct the original inputs of the hash codes.
This thesis addresses the above challenge and makes a number of contributions to representation learning that (i) improve effectiveness of hash codes through more expressive representations and a more effective similarity measure than the current state of the art, namely the Hamming distance, and (ii) improve efficiency of hash codes by learning representations that are especially suited to the choice of search method. The contributions are empirically validated on several tasks related to similarity search and recommendation.


\clearpage

\phantomsection
\addcontentsline{toc}{chapter}{Dansk Resumé}\section*{Dansk Resumé}
Hvordan data repræsenteres og operationaliseres er afgørende for opbygningen af effektive beregningsmodeller. En almindelig tilgang er at repræsentere dataobjekter som binære vektorer, betegnet hash-koder, der kræver lidt lagerplads og muliggør effektiv similaritetssøgning gennem direkte indeksering i en hash-tabel eller gennem similaritetsberegninger i et passende rum.
På grund af den begrænsede ekspressibilitet af hash-koder, sammenlignet med flydende tal repræsentationer, så er en essentiel udfordring hvordan man genererer hash-koder, der repræsenterer semantisk indhold, eller latente egenskaber, godt ved hjælp af et lille antal bits, samtidig med at man sørger for, at hash-koder distribueres på en måde der ikke reducerer deres søgeeffektivitet.
State of the art metoder bruger repræsentationslæring til generering af sådanne hash-koder med fokus på neurale autoencoderarkitekturer, hvor semantik er kodet i hash-koder ved at lære at rekonstruere de originale input af hash-koderne.
Denne afhandling adresserer ovennævnte udfordring og bidrager med en række nye metoder til repræsentationslæring, der (i) forbedrer effektiviteten af hash-koder gennem mere ekspressive repræsentationer og et mere effektivt afstandsmål end den nuværende kendte teknik, nemlig Hamming distancen, og (ii) forbedre søgeeffektiviteten af hash-koder ved at lære repræsentationer, der er særligt velegnede til valget af søgemetode. Bidragene valideres empirisk på flere problemer relateret til similaritetssøgning og recommendation.

\clearpage

\phantomsection
\addcontentsline{toc}{chapter}{Acknowledgements}\section*{Acknowledgements}
My PhD has been a wonderful academic and personal journey that would not have been possible without the support from family, friends, and colleagues. First, I would like to thank my academic advisors Christina, Jakob, and Stephen from whom I have without a doubt learned many valuable lessons. I have especially enjoyed my weekly meetings with Christina and Jakob, where we have had many insightful and entertaining discussions. I would also like to thank my office mates over the years, Dongsheng, Lucas, Niklas, and Stephan, for providing a great work environment with many joyous moments. I was fortunate enough to be able to do an internship at Spotify Research with many fantastic people, where I would especially like to thank Brian, Federico, Lucas, Mounia, and Rishabh. I had a great summer in London! I would also like to thank Benjamin and Rastin from Danmarks Nationalbank, and Esben and Jacob from the Municipality of Copenhagen, it was great discussing and collaborating with you. Next, I would like to thank all my friends (including many of the people mentioned above!) and family, especially David and Mikkel, who have been a steady source of support during the years and made sure I could always see the bigger picture. Lastly, and most importantly, I would like to deeply thank my twin brother Christian for always being there for me and being a great source of inspiration, food, and snacks. 
\clearpage

\tableofcontents
\clearpage
\pagenumbering{arabic}

\chapter{Introduction} 
Similarity search and recommendation are two types of Information Retrieval (IR) tasks used for supporting users in filtering and exploring data, such as searching for interesting books or movies.
To build solutions for such tasks, learning representations of information objects, such as words, documents, and items, is a core challenge in IR, with the aim of capturing key semantics and properties for retrieving the most relevant information. 

Traditional IR text representations have focused on bag-of-words (BoW) representations with a fixed vocabulary (i.e., a limited set of words), based on various term weighting approaches, such as term-frequency \cite{salton1988term}, tf-idf \cite{salton1988term}, BM25 \cite{robertson1995okapi}, and language models \cite{ponte1998language} (see \cite{blanco2012graph,lioma2009part} for an overview on term weighting). 
In BoW vector representations, each dimension corresponds to a specific vocabulary word, which results in highly sparse vectors as the only non-zero dimensions are the vocabulary words occurring in a given text. Due to the sparsity, BoW representations are very efficient to use in practice through fast indexing structures for retrieval.
However, such BoW representations suffer from the problem of \emph{vocabulary mismatch} \cite{furnas1987vocabulary}. Vocabulary mismatch refers to the problem when different words or phrases are used to express the same meaning, which occurs for BoW representations as each word is only mapped to a single dimension \cite{zhao2010term}. 

One way to address the vocabulary mismatch problem is through \emph{distributional semantics}, which is based on the distributional hypothesis stating that words occurring in the same context tend to have similar meanings \cite{harris1954distributional}. In distributional semantics, text representations are typically represented as dense vectors of significantly fewer dimensions than the vocabulary size, where each word is distributed across all dimensions. Such vectors are typically denoted as embeddings, as they are obtained by embedding high dimensional data (or complex data objects generally) into a relatively low dimensional space in such a way that semantically similar data is placed close together in the embedding space\footnote{Note that many types of data can be effectively embedded, not just words and documents, such that data objects close together in the embedding space correspond to objects with similar properties.}.
Early work on distributional semantics for representing documents was based on computing a matrix decomposition of the document-term co-occurrence matrix, e.g., using the seminal Latent Semantic Indexing (LSI) approach \cite{deerwester1990indexing}, such that similar documents are represented as similar embeddings. However, computing the decomposition of the co-occurrence matrix is computationally expensive on large-scale data due to the runtime complexity of the singular value decomposition typically used \cite{schutze2008introduction}.
In a less computationally expensive way, embeddings have also been constructed on a word level, where methods such as word2vec~\cite{word2vec} and GloVe~\cite{pennington2014glove} learn word embeddings based on word-word co-occurrences. Specifically, these methods aim at learning word embeddings that are able to predict a word given its context, or the context given a word, where the context is represented by $n$ words before and after a given word (denoted as $n$-length context windows). The use of context windows can be seen in contrast to using all the words in a document as done by LSI. To create document embeddings, embeddings of the words within a document can be aggregated, e.g., by taking an average \cite{lev2015defense}, a frequency-weighted average \cite{Arora2017ASB}, or through max-pooling \cite{Zhelezniak2019DontSF}. 
However, while such embeddings provide an effective encoding of document semantics, the cost associated with similarity computations (e.g., using the inner product), as well as the amount of storage required, is nontrivial and may hinder use in very large-scale settings. We will next describe a type of representation that supports faster similarity computations while simultaneously requiring significantly less storage. 

\subsection*{Hash Code Representations}
An efficient approach for representing documents (or more generally \emph{data objects}) is through short binary vector representations denoted as \emph{hash codes}. 
In contrast to real-valued embeddings, hash codes may require only a few bytes of storage, and support fast similarity computations through the Hamming distance, which can be computed in two CPU instructions counting the number of differing bits between two hash codes. Additionally, hash codes can be used for direct indexing into a hash table for finding similar documents with exact hash code matches, or by making appropriate bit flips for directly finding documents up to a specific Hamming distance to the reference (i.e., query) hash code.
Locality-Sensitive Hashing (LSH) \cite{indyk1998approximate,gionis1999similarity} is a widely known family of hashing techniques, where highly similar (i.e., near-identical) documents have a high probability of hashing to the same hash code.
However, LSH uses data-independent hashing functions for mapping similar documents to similar hash codes, such as random projections. Data-independent hashing functions have the limitation of not utilizing any training data in their construction, which limits their ability to encode semantics. For this reason, LSH techniques are less suited for computing the semantic similarity between documents. 

In contrast, semantic hashing~\cite{salakhutdinov2009semantic} methods are data-dependent and learn a mapping for transforming documents into semantically-aware hash codes, such that semantically similar documents have a short Hamming distance between their associated hash codes. 
Some of the first work on semantic hashing utilized quantized (i.e., binary) versions of familiar techniques, such as LSI \cite{zhang2010laplacian} and spectral clustering \cite{spectral-hashing-nips}, and through two step approaches of first finding the optimal encoding and then learning to predict this as a classification problem \cite{zhang2010self,spectral-hashing-nips}. 
Recent work has focused on neural network approaches, specifically variational autoencoder architectures \cite{chaidaroon2017variational,chaidaroon2018deep,nash2018,dong-etal-2019-document,DBLP:conf/sigir/hansen-sigir2020-pairwisehash,DBLP:conf/sigir/hansen-sigir19-semhash,DBLP:conf/www/hansen-www2021-mish}. These models are trained with the objective of reconstructing the input representation (typically a tf-idf BoW representation) from the latent representation of the most internal layer in the autoencoder, which corresponds to the hash code. In this way, such models learn an encoding that captures the key semantics within the hash code, as it otherwise would be unable to reconstruct it well. Learned hash codes have been found to be effective not only in document similarity search, but also for representing users and items in recommender systems \cite{karatzoglou2010collaborative,zhou2012learning,zhang2014preference,Lian:2017:DCM:3097983.3098008,Zhang:2018:DDL:3159652.3159688,DBLP:conf/sigir/hansen-sigir2020-coldstart,DBLP:conf/www/hansen-www2021-phd}. 
In this setting, user representations act as query objects for which the most relevant items should be retrieved. This is analogous to the document similarity search setting where documents act as both queries and items to be retrieved.

This thesis addresses the challenges of generating efficient hash codes that well capture semantic content, or latent properties, using a small number of bits. We make a number of contributions to representation learning that (i) improve effectiveness of hash codes through more expressive representations and a more effective similarity measure than the current state of the art, namely the Hamming distance, and (ii) improve efficiency of hash codes by learning representations that are especially suited to the choice of search method. The contributions are empirically validated on several tasks related to similarity search and recommendation.

\section{Research Outline}\label{s:research-outline}


This thesis is composed of five chapters, where each chapter is a published article (see Section \ref{s:publication} for the list of publications). These five chapters are grouped into the application areas of document similarity search (Section \ref{ss:doc-sim-search}) and recommendation (Section \ref{ss:recommendation}).

This section provides an outline of each chapter of the thesis, where we state the main research questions, provide background information, and overview the main findings and observations.

\subsection{Document Similarity Search} \label{ss:doc-sim-search}
The task of document similarity search consists of searching a set of document objects by a given query object, such that those document objects most similar to the query are retrieved. 
In the setting of document similarity search using hash codes, similarity search is usually considered either a radius search or a k-nearest neighbour (kNN) search. In a radius search, all hash codes with a specified maximum distance to the query hash code are to be found, whereas for kNN the radius is incrementally increased until the k$^\textrm{th}$ hash code has a distance equal to the search radius. In our work on document similarity search using semantic hashing, we focus on kNN search as per related work \cite{chaidaroon2017variational,chaidaroon2018deep,nash2018,dong-etal-2019-document}, but note that the work can trivially be adapted for radius search.

\subsubsection{Chapter 2. Unsupervised Neural Generative Semantic Hashing}
Existing approaches for semantic hashing are typically unsupervised as document labels are most often not available for large data collections of a size where fast similarity search using hash codes is particularly useful.
The approaches have evolved into three main groups. Firstly, being based on two-step procedures of first learning the optimal hash code encoding (based on low Hamming distances between semantically similar documents) and then learning to predict this to handle unseen documents \cite{spectral-hashing-nips,zhang2010self}. 
Secondly, variational autoencoder models with an input reconstruction loss using a post-processing quantization step for obtaining the hash codes \cite{chaidaroon2017variational,chaidaroon2018deep}. The quantization step consists of rounding each dimension based on its median value, such that the k$^\textrm{th}$ bit of the hash code is 1 if the k$^\textrm{th}$ dimension of the learned (real-valued) vector is larger than the median, and vice versa for setting a bit to 0.
Thirdly, variational autoencoders using Bernoulli priors for learning hash codes in an end-to-end fashion \cite{nash2018}, which improves hash code effectiveness due to reducing the quantization error. However, none of these approaches directly model the goal of ranking documents by their similarity to the query document, i.e., for maximizing the precision of a kNN search, but rather implicitly assume that the learned hash codes will enable this by focusing on encoding the document semantics well. This leads us to the first research question:
\begin{description}
\item[RQ1] To what extent can ranking become an organic part of learning semantic
hash codes?
\end{description}
To answer this question, we first need to obtain relevance labels between a query document and the remaining documents. As we work in an unsupervised setting, this cannot be based on provided document tags or labels, but rather we take a weakly supervised direction by using an existing unsupervised semantic hashing approach for obtaining approximate top-K rankings of each document. Based on this, we extract ranking triplets for training, such that each sample consists of query document, a similar document, and a dissimilar document. With such ranking triplets, the aim is to learn hash codes that are better able to separate similar and dissimilar documents in the Hamming space.
We propose Ranking-based Semantic Hashing (RBSH), a variational autoencoder with a traditional input reconstruction objective trained jointly with a hinge loss on the ranking triplets as the ranking objective.
We find that the ranking objective has a beneficial regularizing effect, as the hash codes, especially short hash codes down to 8 bits, otherwise have a tendency to cluster and not sufficiently utilize the space, which reduces the kNN effectiveness. We experimentally find that RBSH hash codes significantly outperform state-of-the-art approaches, and most importantly, yield state-of-the-art performance while using 2-4x fewer bits than existing approaches.

\subsubsection{Chapter 3. Unsupervised Semantic Hashing with Pairwise Reconstruction}
The findings from \textbf{RQ1} suggest that incorporating neighbourhood information, in our case in the form of optimizing a joint ranking objective, can improve the generalization of the learned hash codes as observed by higher kNN effectiveness. This observation is also noted in early work \cite{spectral-hashing-nips,zhang2010self}, where the cosine similarity between documents in their original space is used for constructing pairwise weights for learning to minimize a weighted Hamming distance between hash code pairs. In this way, hash codes from similar documents are forced to have a small Hamming distance, whereas dissimilar documents with a small or negative weight are ignored or forced further apart in the Hamming space. In more recent work, a variational autoencoder model, denoted NbrReg, with two reconstruction objectives (represented as two different decoders in the autoencoder) is proposed \cite{chaidaroon2018deep}. The first reconstruction objective is the typical input reconstruction, whereas the second aims at reconstructing all unique words occurring in a local neighbourhood around the input document (i.e., an aggregated neighbourhood document). Based on this line of work, we ask the following research question:
\begin{description}
\item[RQ2] To what extent can local semantic neighbourhoods be incorporated as an organic part of learning semantic hash codes?
\end{description}
We answer this question by proposing an extension to the input reconstruction objective shared among the variational autoencoder approaches \cite{chaidaroon2017variational,chaidaroon2018deep,nash2018,DBLP:conf/sigir/hansen-sigir19-semhash}. Similarly to our previously proposed RBSH (\textbf{RQ1}), we use an existing unsupervised semantic hashing approach to retrieve a set of similar documents to each query document. Using this set, we construct training pairs and train a variational autoencoder, named PairRec, to be able to reconstruct the query document from both hash codes (i.e., a pairwise reconstruction). In this way, more general hash codes are learned that not only encode their own semantics, but also those of similar documents, which we experimentally show to be more effective than existing state-of-the-art approaches. 
In contrast to NbrReg \cite{chaidaroon2018deep}, we consider pairs of real documents, as opposed to the aggregated neighbourhood document used by NbrReg, which we argue may introduce a semantic shift as such documents cannot be encountered during retrieval.
Furthermore, our approach uses only a single decoder, as the two decoders used by NbrReg increase the overall decoding complexity, which may be difficult to capture using the simple Hamming distance and hence unnecessarily reduce the kNN effectiveness.

\subsubsection{Chapter 4. Unsupervised Multi-Index Semantic Hashing}
\textbf{RQ1} and \textbf{RQ2} address ways for learning more expressive hash codes that improve kNN effectiveness. However, both our and prior work have generally assumed that the hash codes will be efficient to use in practice, but not considered any concrete search method to explore whether any efficiency differences exist between the approaches. While real-time brute-force linear scans are possible using the Hamming distance on large-scale data \cite{shan2018recurrent}, significantly faster alternatives exist enabling sub-linear search time. One such alternative is multi-index hashing \cite{norouzi2012fast,norouzi2013fast}, a method for performing exact radius and kNN search in the Hamming space. While hash codes can be used as direct addresses into a hash table, the number of such lookups scales exponentially with the radius (where the number of bits is the base), which can become infeasible for long hash codes or even moderate radii. To fix this, multi-index hashing splits each hash code into $m$ disjoint substrings, and utilizes the pigeonhole principle to determine that if two hash codes are within radius $r$, then at least one substring exists where the Hamming distance between the substrings is at most $\lfloor{\frac{r}{m}}\rfloor{}$. Based on this, multi-index hashing builds a candidate set, significantly smaller than the entire set of hash codes, from which a linear scan is performed to determine the exact Hamming distances. This leads to our next research question:
\begin{description}
\item[RQ3] To what extent can key properties for multi-index hashing become an organic part of learning semantic hash codes?
\end{description}
To answer this question, we first identify two key properties for high multi-index efficiency. The first one is to reduce the number of documents per hash table lookup by reducing the number of false-positive candidates, i.e., those with a small substring distance, but high overall hash code distance. The second one is to distribute the hash codes sufficiently such that the distance to the k$^\textrm{th}$ document is kept low to limit the exponential growth. We operationalize these properties into model-agnostic training objectives for training hash codes specifically designed for multi-index hashing in an end-to-end fashion, which we denote as Multi-Index Semantic Hashing (MISH). We experimentally find that state-of-the-art baselines are upwards of 33\% slower than hash codes generated by MISH without being more effective.

\subsection{Recommendation} \label{ss:recommendation}
Recommender systems are trained on data from user-item interactions, e.g., clicks or ratings, with the objective of learning to estimate user preferences to provide relevant recommendations. When users and items are represented as embeddings, such as hash codes, the recommendation task is highly similar to that of similarity search (Section \ref{ss:doc-sim-search}), but with the modification that the user representation acts as the query in contrast to using an item as a query. In the following chapters, we focus on two different recommendation settings, depending on the availability of content information, but they have the commonality of generating user and item hash codes to be used for recommendation.

\subsubsection{Chapter 5. Content-aware Neural Hashing for Cold-start Recommendation}
Recommendation approaches based on collaborative filtering, content-based filtering, and their combinations have been well studied and shown to work well in practice \cite{adomavicius2005toward,shi2014collaborative}. Collaborative filtering learns user and item embeddings based on user-item interactions, such as implicit feedback (e.g., clicks) or explicit feedback (e.g., ratings). However, when new items appear (denoted \emph{cold-start} items) no previous user-item interactions exist, hence collaborative filtering approaches are unable to learn such embeddings. Content-based filtering solves the cold-start problem by using item content information (or other item features) for recommending items similar to items a user has previously enjoyed. For efficiency reasons, a number of hashing-based approaches have also been proposed for the recommendation domain. These approaches have primarily focused on the collaborative filtering setting \cite{karatzoglou2010collaborative,zhang2014preference,zhou2012learning,liu2019compositional,zhang2016discrete}, but less so on content-aware approaches addressing the cold-start problem. Existing content-aware hashing approaches, DDL \cite{zhang2018discrete} and DCMF \cite{lian2017discrete}, learn to generate user and item hash codes for use in both standard and cold-start settings, however they both share the problem of generating item hash codes differently depending on whether the item is considered cold-start or not. Specifically, they both learn user and item hash codes in a typical collaborative filtering setting, but simultaneously learn separate item hash codes based on their content information, such that the distance between the two types of item hash codes is minimized. We argue that this is unnecessary and may limit generalizability, which leads to the next research question: 
\begin{description}
\item[RQ4] How can item content information be used for generating hash codes in the same way for both standard and cold-start items to improve recommendation effectiveness?
\end{description}
To answer this question, we propose NeuHash-CF, a joint hashing model for generating user and item hash codes robust to the cold-start problem. Inspired by semantic hashing, we use the basic architecture of our previous work \cite{DBLP:conf/sigir/hansen-sigir19-semhash,DBLP:conf/sigir/hansen-sigir2020-pairwisehash} to construct an item component that learns hash codes based on content information alone. In contrast, the user component is based solely on a user ID as in the typical collaborative filtering setting, and the model is jointly optimized by learning to reconstruct the logged user-item ratings. Since the item hash codes are generated entirely based on content information, they are by default robust to the cold-start problem as long as new items share some similarity to some of the existing items. We experimentally evaluate NeuHash-CF hash codes against state-of-the-art baselines, including collaborative filtering and content-aware approaches, where we observe significant improvements in both cold-start and standard recommendation settings.

\subsubsection{Chapter 6. Projected Hamming Dissimilarity for Bit-Level Importance Coding in Collaborative Filtering}
In hashing-based learning, such as for documents (\textbf{RQ1-3}) or users and items (\textbf{RQ4}), the default way of measuring the similarity between two hash codes is through the Hamming distance, which is computed as the summation of the XOR operation between two hash codes. However, by definition, the Hamming distance weighs each bit equally, which could be problematic when the importance of each bit's underlying properties differ depending on the query. For example, in a collaborative filtering setting the user hash code represents the query, and depending on the user's historic item interactions, it may be possible to infer that certain underlying properties are more important for the item ranking. 
While approaches have been proposed for assigning real-valued weights to certain substrings of bits in a hash code \cite{liu2019compositional,fan2013learning}, such a weighting has the problem of making the core similarity computation (e.g., Hamming distance) significantly slower, which limits its usage in large-scale settings where hashing-based solutions are most needed. This leads to the next research question:
\begin{description}
\item[RQ5] How can a similarity measure support hash code importance weighting without reducing efficiency?
\end{description}
To answer this question, we consider the collaborative filtering setting of user and item hash codes. We derive a new way of measuring the dissimilarity between a user (query) and item hash code with binary weighting of each dimension, corresponding to disabling specific bits. To this end, we consider a general definition of dissimilarity defined as the norm of the difference between the user representation and the item representation projected onto the user, which in the Euclidean space corresponds to the well known cosine distance. We show that when working in the Hamming space, this results in a novel projected Hamming dissimilarity, which by choice of projection, effectively allows an importance weighting of the item hash codes through the user hash code. Specifically, if the possible bit values are $\{-1,1\}$, then all bit dimensions with $-1$ in the user hash code are ignored in the dissimilarity computation across all items, corresponding to a binary importance weighting of each bit dimension. We experimentally show that hash codes optimized for the projected Hamming dissimilarity lead to large gains in recommendation effectiveness compared to baselines using the Hamming distance, while requiring no additional storage and no reduction in efficiency compared to computing the Hamming distance.

\section{Summary of Contributions}
This thesis addresses challenges related to representation learning for enabling more efficient and effective similarity search and recommendation. The contributions focus on improving the representational power of learned hash codes, as well as learning to construct them in a way that enables more efficient similarity search. 

\begin{itemize}
\item The first contribution is a method for retrieving semantically relevant documents based on learned hash codes, which are constructed in a way that encodes both document semantics and ranking information relative to other hash codes. The ranking part ensures that the hash codes are directly designed for being ranked, rather than only encoding semantics, which improves retrieval effectiveness.

\item The second contribution is an improved method of learning document hash codes, based on introducing local semantic neighbourhood information into the learning process. Specifically, hash codes from pairs of semantically similar documents are trained such that both hash codes can reconstruct the same original document, which effectively ensures that semantically similar documents are placed close together in the learned hash code space.

\item The third contribution is a method for learning hash codes that enables more efficient retrieval by directly optimizing the hash code construction towards the search approach being used. In contrast to prior work that applies post-processing steps after training for improving efficiency, we find that learning it in a direct end-to-end trainable way enables even larger efficiency gains without compromising effectiveness.

\item The fourth contribution is a method for learning user and item hash codes for recommendation, while being robust to the problem of cold-start items. This is achieved by using content information for generating item hash codes in a unified way for both standard and cold-start items, rather than distinguishing between them as done by prior work, which improves recommendation effectiveness.

\item The fifth contribution is a new dissimilarity measure for comparing two hash codes, which enables a binary weighting of the hash codes, corresponding to disabling the bits deemed unimportant by one of the hash codes. Hash codes optimized for this dissimilarity, rather than the current state of the art, namely the Hamming distance, results in higher effectiveness in collaborative filtering experiments, while requiring no additional storage and no computational overhead compared to the Hamming distance. 
\end{itemize}

\section{Future Work}
Our contributions towards more efficient and effective representations for similarity search and recommendation naturally have their limitations and possible directions of future work. Below we first discuss specific directions of future work for research presented in this thesis, and follow with a discussion of more general directions of future work.

\subsection*{New input representations}
All text hashing approaches proposed in this thesis, as well as the baselines used for comparison, use a tf-idf weighted bag-of-words vector as the document representation. As state-of-the-art approaches all use variational autoencoders, the motivation for using a fixed input vector is for using it in reconstruction when decoding the hash code as part of training the autoencoder. As an alternative, Doan and Reddy \cite{doan2020efficient} recently explored two approaches using recurrent and convolutional neural networks, with the objective of being able to reconstruct the word embeddings of the input. They empirically found this to be more effective than the (tf-idf vector) baselines, but they did not compare against more recent state-of-the-art approaches, so the exact improvement is uncertain. However, their work does provide a first step towards exploring more expressive representations than a tf-idf vector, which is a natural next step for our work as well. 

\subsection*{Exploring the projected Hamming dissimilarity and other functions in new settings}
We proposed the projected Hamming dissimilarity as a way to compute weighted dissimilarities in the asymmetrical user-item setting, where the user hash code acts as a query used for searching among the item hash codes. While our work showed large effectiveness gains, a next step would be exploring it (as well as other potential functions) in a symmetrical item-item setting, such as document similarity search. A potential obstacle is that each item also acts as a query for similarity search, which may be problematic as the weighting in the projected Hamming dissimilarity is defined through the query, hence its dual purpose could hinder learning a good representation. However, a first step would be to simply learn two hash codes per item, one for the purpose of a query, and one for the normal purpose of an item. Note that this does not necessarily lead to increased storage requirements, as the similarity search often is performed on new (unseen) documents, which would require an encoding step for obtaining the hash code in any case. 

\subsection*{Hashing-based representations in dynamic settings}
This thesis has considered application areas where item (and user) hash codes could be generated without considering the need for any updates. While the hash code models should be retrained regularly, and new hash codes should be generated when new data arrives, this has not been a focus in our work. To this end, it would be interesting to investigate the effectiveness of hashing-based approaches for more dynamic settings, such as context-aware or feedback-aware recommender systems \cite{dai2016deep,beutel2018latent,wu2017recurrent,DBLP:conf/recsys/hansen-recsys-spotify}, where the user representation should change often (e.g., after every few interactions) to better represent the (contextual or temporal) needs of the user. While more expressive real-valued representations might be needed for the final item ranking, the hash codes could prove useful for efficiently generating highly relevant candidate pools.

\subsection*{Representation interpretability}
Interpretability is a core challenge for representation learning, as well as for machine learning as a whole. Investigating the relation between interpretable concepts and how embeddings are constructed, whether being real-valued or binary, is important for understanding what a model has learned, which may improve the trust we as humans assign to such models for use in real-world applications. 
For real-valued embeddings, one way to improve interpretability is through inducing sparsity when constructing the embeddings \cite{subramanian2018spine}, such that the embedding of the document (or another type of object) can be described by a few non-zero dimensions, corresponding to some latent topics. Each topic, or combination of topics, would then be described by a set of documents from which a common theme could be extracted. However, for hash codes, this type of sparsity is not as straight-forward to induce because hash codes are unable to represent the equivalent of a real-valued zero, as both possible binary values impact the distance computation. As little work has been done in hash code interpretability, it would be worth pursuing that direction in the future.

\subsection*{Representation pretraining}
Learning representations that generalize well is another important challenge, but doing so may require large and varied amounts of appropriate training data. One way to solve this problem is through model pretraining, with one of the most successful examples being word embedding models (e.g., word2vec~\cite{word2vec} and BERT \cite{devlin2019bert}). These models are trained in a unsupervised fashion on massive collections of text, from which general language understanding is learned, such that the models later can be fine-tuned (or even applied directly) to downstream tasks. Utilizing, and further exploring, this idea of pretraining is highly attractive, as it allows new models to exploit the language understanding obtained through the pretraining for improving their generalization. Interestingly, such pretrained models have not yet been explored for the task of learning semantic hash codes for documents, neither through using existing word embedding models or (pre)training hash code models on the same massive collections of text as the original word embedding models.

\clearpage
\section{List of Publications}\label{s:publication}
The following publications are included as chapters of this thesis ($^*$ denotes equal contribution):
\begin{description}
    \item[Chapter \ref{chapter2}] Casper Hansen, Christian Hansen, Jakob Grue Simonsen, Stephen Alstrup, Christina Lioma (2019). Unsupervised Neural Generative Semantic Hashing. In SIGIR, pages 735-744. \cite{DBLP:conf/sigir/hansen-sigir19-semhash}.
    
    \item[Chapter \ref{chapter3}] Casper Hansen$^*$, Christian Hansen$^*$, Jakob Grue Simonsen, Stephen Alstrup, Christina Lioma (2020). Unsupervised Semantic Hashing with Pairwise Reconstruction. In SIGIR, pages 2009-2012. \cite{DBLP:conf/sigir/hansen-sigir2020-pairwisehash}.
    
    \item[Chapter \ref{chapter4}] Christian Hansen$^*$, Casper Hansen$^*$, Jakob Grue Simonsen, Stephen Alstrup, Christina Lioma (2021). Unsupervised Multi-Index Semantic Hashing. In WWW, pages 2879-2889. \cite{DBLP:conf/www/hansen-www2021-mish}.
    
    \item[Chapter \ref{chapter5}] Casper Hansen$^*$, Christian Hansen$^*$, Jakob Grue Simonsen, Stephen Alstrup, Christina Lioma (2020). Content-aware Neural Hashing for Cold-start Recommendation. In SIGIR, pages 971-980. \cite{DBLP:conf/sigir/hansen-sigir2020-coldstart}.
    
    \item[Chapter \ref{chapter6}] Christian Hansen$^*$, Casper Hansen$^*$, Jakob Grue Simonsen, Christina Lioma (2021). Projected Hamming Dissimilarity for Bit-Level Importance Coding in Collaborative Filtering. In WWW, pages 261-269. \cite{DBLP:conf/www/hansen-www2021-phd}.
\end{description}
Furthermore, publications not part of this thesis have been written during the PhD studies on the following topics:
(non hashing-based) recommender systems \cite{wsdm-spotify,DBLP:conf/recsys/hansen-recsys-spotify,DBLP:conf/www/hansen-wsdmcup-2019}, computational approaches for COVID-19 prediction and retrieval \cite{jimenez2021developing,trec-covid-lucas}, fact-checking \cite{hansen-etal-2021-automatic,DBLP:conf/clef/hansen-clef-2020,DBLP:conf/sigir/sigir2020-eyetracking,augenstein-etal-2019-multifc,hansen2019copenhagen,DBLP:conf/www/hansen-www2019-factcheck,hansen2018copenhagen}, (non hashing-based) text representation and classification \cite{dongsheng-ecir2021,DBLP:conf/sigir/hansen-sigir2019-contextprop,DBLP:conf/iclr/hansen-iclr,MATIN2019199}, smart city analytics \cite{DBLP:conf/cikm/hansen-cikm2017}, and educational data mining \cite{DBLP:conf/edm/hansen-edm-2019,ecel-2017,DBLP:conf/edm/edm2017}. These publications are listed below in reverse chronological order:
\begin{enumerate}
    \item Casper Hansen$^*$, Christian Hansen$^*$, Lucas Chaves Lima (2021). Automatic Fake News Detection: Are Models Learning to Reason? In ACL, pages 80-86. \cite{hansen-etal-2021-automatic}. 
    \item Dongsheng Wang$^*$, Casper Hansen$^*$, Lucas Chaves Lima, Christian Hansen, Maria Maistro, Jakob Grue Simonsen, Christina Lioma (2021). Multi-Head Self-Attention with Role-Guided Masks. In ECIR, pages 432-439. \cite{dongsheng-ecir2021}.
    \item Christian Hansen, Rishabh Mehrotra, Casper Hansen, Brian Brost, Lucas Maystre, Mounia Lalmas (2021). Shifting Consumption towards Diverse Content on Music Streaming Platforms. In WSDM, pages 238-246. \cite{wsdm-spotify}.
    \item Espen Jimenez Solem, Tonny Studsgaard Petersen, Casper Hansen, Christian Hansen, et al. (2021). Developing and Validating COVID-19 Adverse Outcome Risk Prediction Models from a Bi-national European Cohort of 5594 Patients. In Scientific Reports 11 (1), pages 1-12. \cite{jimenez2021developing}.
    \item Casper Hansen, Christian Hansen, Lucas Maystre, Rishabh Mehrotra, Brian Brost, Federico Tomasi, Mounia Lalmas (2020). Contextual and Sequential User Embeddings for Large-Scale Music Recommendation. In RecSys, pages 53-62. \cite{DBLP:conf/recsys/hansen-recsys-spotify}.
    \item Lucas Chaves Lima$^*$, Casper Hansen$^*$, Christian Hansen, Dongsheng Wang, Maria Maistro, Birger Larsen, Jakob Grue Simonsen, Christina Lioma (2021). Denmark's Participation in the Search Engine TREC COVID-19 Challenge: Lessons Learned about Searching for Precise Biomedical Scientific Information on COVID-19. In TREC COVID-19 Challenge. \cite{trec-covid-lucas}.
    \item Casper Hansen, Christian Hansen, Jakob Grue Simonsen, Christina Lioma (2020). Fact Check-Worthiness Detection with Contrastive Ranking. In CLEF, pages 124-130. \cite{DBLP:conf/clef/hansen-clef-2020}.
    \item Christian Hansen, Casper Hansen, Jakob Grue Simonsen, Birger Larsen, Stephen Alstrup, Christina Lioma (2020). Factuality Checking in News Headlines with Eye Tracking. In SIGIR, pages 2013-2016. \cite{DBLP:conf/sigir/sigir2020-eyetracking}.
    \item Isabelle Augenstein, Christina Lioma, Dongsheng Wang, Lucas Chaves Lima, Casper Hansen, Christian Hansen, Jakob Grue Simonsen (2019). MultiFC: A Real-World Multi-Domain Dataset for Evidence-Based Fact Checking of Claims. In EMNLP, pages 4685-4697. \cite{augenstein-etal-2019-multifc}.
    \item Casper Hansen, Christian Hansen, Jakob Grue Simonsen, Christina Lioma (2019). Neural Weakly Supervised Fact Check-Worthiness Detection with Contrastive Sampling-Based Ranking Loss. In CLEF-2019 Fact Checking Lab. \cite{hansen2019copenhagen}.
    \item Casper Hansen, Christian Hansen, Stephen Alstrup, Jakob Grue Simonsen, Christina Lioma (2019). Contextually Propagated Term Weights for Document Representation. In SIGIR, pages 897-900. \cite{DBLP:conf/sigir/hansen-sigir2019-contextprop}.
    \item Christian Hansen, Casper Hansen, Stephen Alstrup, Christina Lioma (2019). Modelling End-of-Session Actions in Educational Systems. In EDM, pages 306-311. \cite{DBLP:conf/edm/hansen-edm-2019}.
    \item Casper Hansen, Christian Hansen, Stephen Alstrup, Jakob Grue Simonsen, Christina Lioma (2019). Neural Check-Worthiness Ranking with Weak Supervision: Finding Sentences for Fact-Checking. In Companion Proceedings of WWW, pages 994-1000. \cite{DBLP:conf/www/hansen-www2019-factcheck}.
    \item Christian Hansen, Casper Hansen, Stephen Alstrup, Jakob Grue Simonsen, Christina Lioma (2019). Neural Speed Reading with Structural-Jump-LSTM. In ICLR. \cite{DBLP:conf/iclr/hansen-iclr}.
    \item Christian Hansen, Casper Hansen, Jakob Grue Simonsen, Stephen Alstrup, Christina Lioma (2019). Modelling Sequential Music Track Skips Using a Multi-RNN Approach. In WSDM Cup. \cite{DBLP:conf/www/hansen-wsdmcup-2019}.
    \item Rastin Matin, Casper Hansen, Christian Hansen, Pia Mølgaard (2019). Predicting Distresses using Deep Learning of Text Segments in Annual Reports. In Expert Systems With Applications (132), pages 199-208. \cite{MATIN2019199}.
    \item Casper Hansen, Christian Hansen, Jakob Grue Simonsen, Christina Lioma (2018). The Copenhagen Team Participation in the Check-Worthiness Task of the Competition of Automatic Identification and Verification of Claims in Political Debates of the CLEF2018 CheckThat! Lab. In CLEF-2018 Fact Checking Lab. \cite{hansen2018copenhagen}.
    \item Casper Hansen, Christian Hansen, Stephen Alstrup, Christina Lioma (2017). Smart City Analytics: Ensemble-Learned Prediction of Citizen Home Care. In CIKM, pages 2095-2098. \cite{DBLP:conf/cikm/hansen-cikm2017}.
    \item Stephen Alstrup, Casper Hansen, Christian Hansen, Niklas Hjuler, Stephan Lorenzen, Ninh Pham (2017). DABAI: A data driven project for e-Learning in Denmark. In ECEL, pages 18-24. \cite{ecel-2017}.
    \item Christian Hansen, Casper Hansen, Niklas Hjuler, Stephen Alstrup, Christina Lioma (2017). Sequence Modelling For Analysing Student Interaction with Educational Systems. In EDM, pages 232-237. \cite{DBLP:conf/edm/edm2017}.
\end{enumerate}

\chapter{Unsupervised Neural Generative Semantic Hashing} \label{chapter2}
Casper Hansen, Christian Hansen, Jakob Grue Simonsen, Stephen Alstrup, Christina Lioma (2019). Unsupervised Neural Generative Semantic Hashing. In SIGIR, pages 735-744. \cite{DBLP:conf/sigir/hansen-sigir19-semhash}.
\includepdf[pages=-, pagecommand={}]{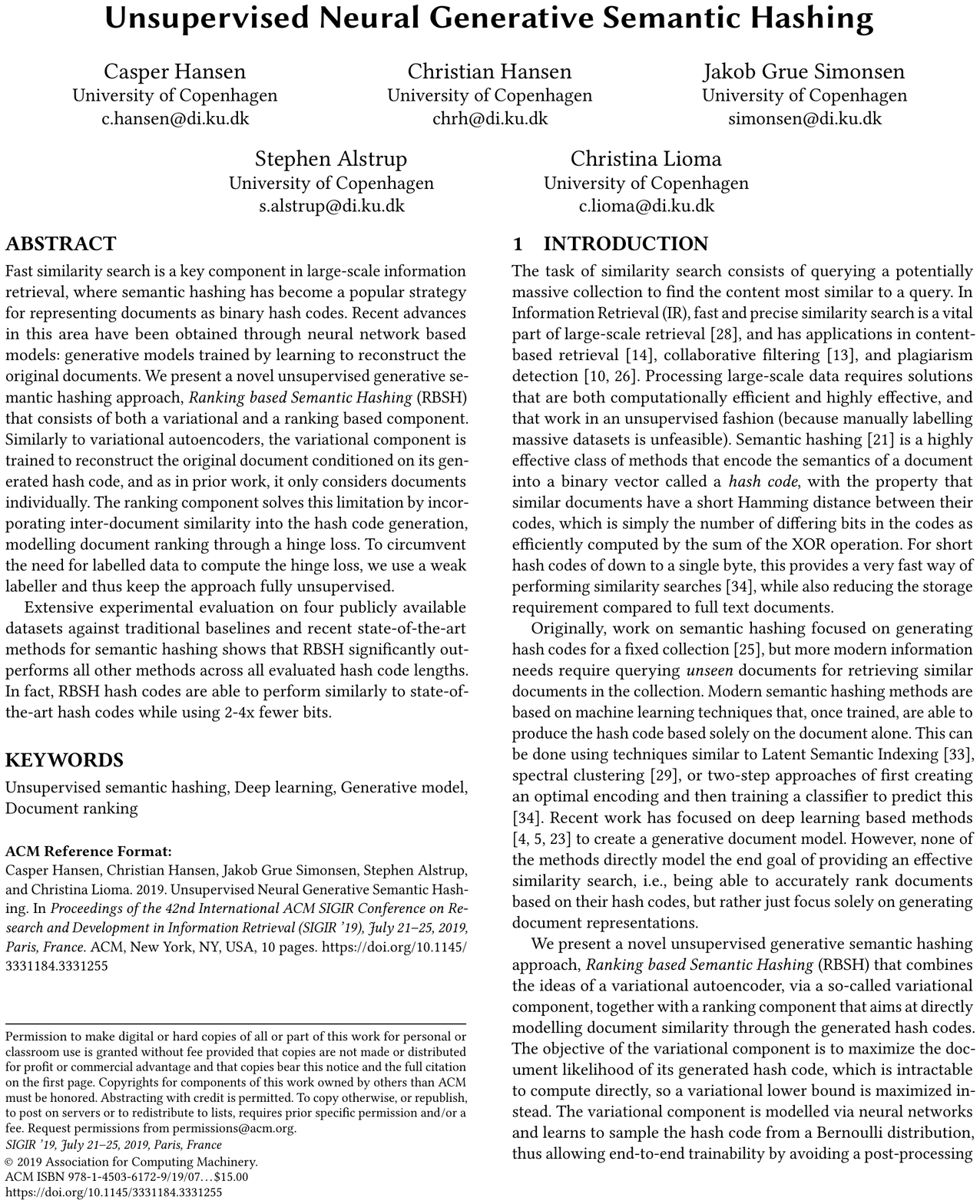}

\chapter{Unsupervised Semantic Hashing with Pairwise Reconstruction} \label{chapter3}
Casper Hansen$^*$, Christian Hansen$^*$, Jakob Grue Simonsen, Stephen Alstrup, Christina Lioma (2020). Unsupervised Semantic Hashing with Pairwise Reconstruction. In SIGIR, pages 2009-2012. \cite{DBLP:conf/sigir/hansen-sigir2020-pairwisehash}. $^*$ denotes equal contribution.
\includepdf[pages=-, pagecommand={}]{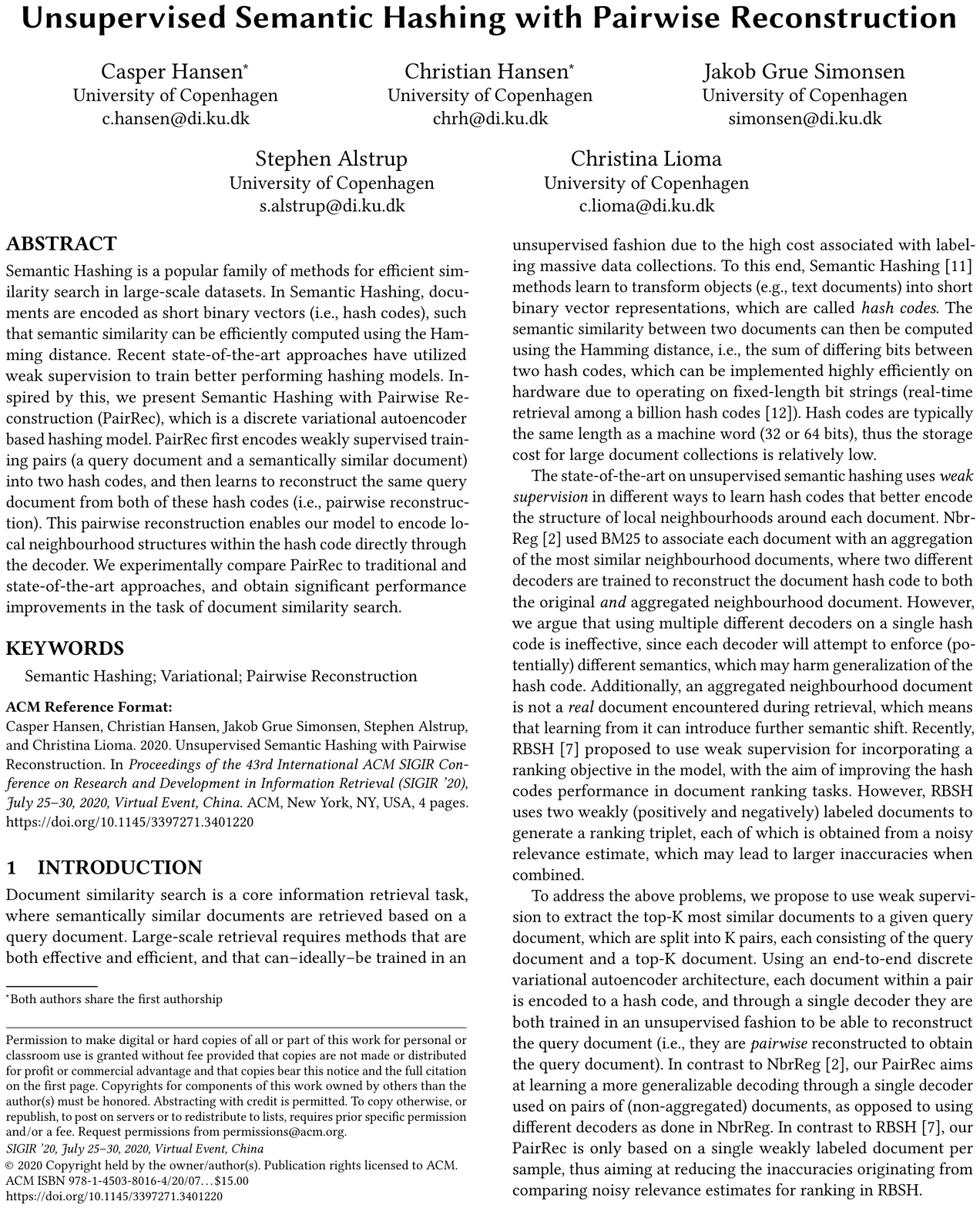}

\chapter{Unsupervised Multi-Index Semantic Hashing} \label{chapter4}
Christian Hansen$^*$, Casper Hansen$^*$, Jakob Grue Simonsen, Stephen Alstrup, Christina Lioma (2021). Unsupervised Multi-Index Semantic Hashing. In WWW, in press. \cite{DBLP:conf/www/hansen-www2021-mish}. $^*$ denotes equal contribution.
\includepdf[pages=-, pagecommand={}]{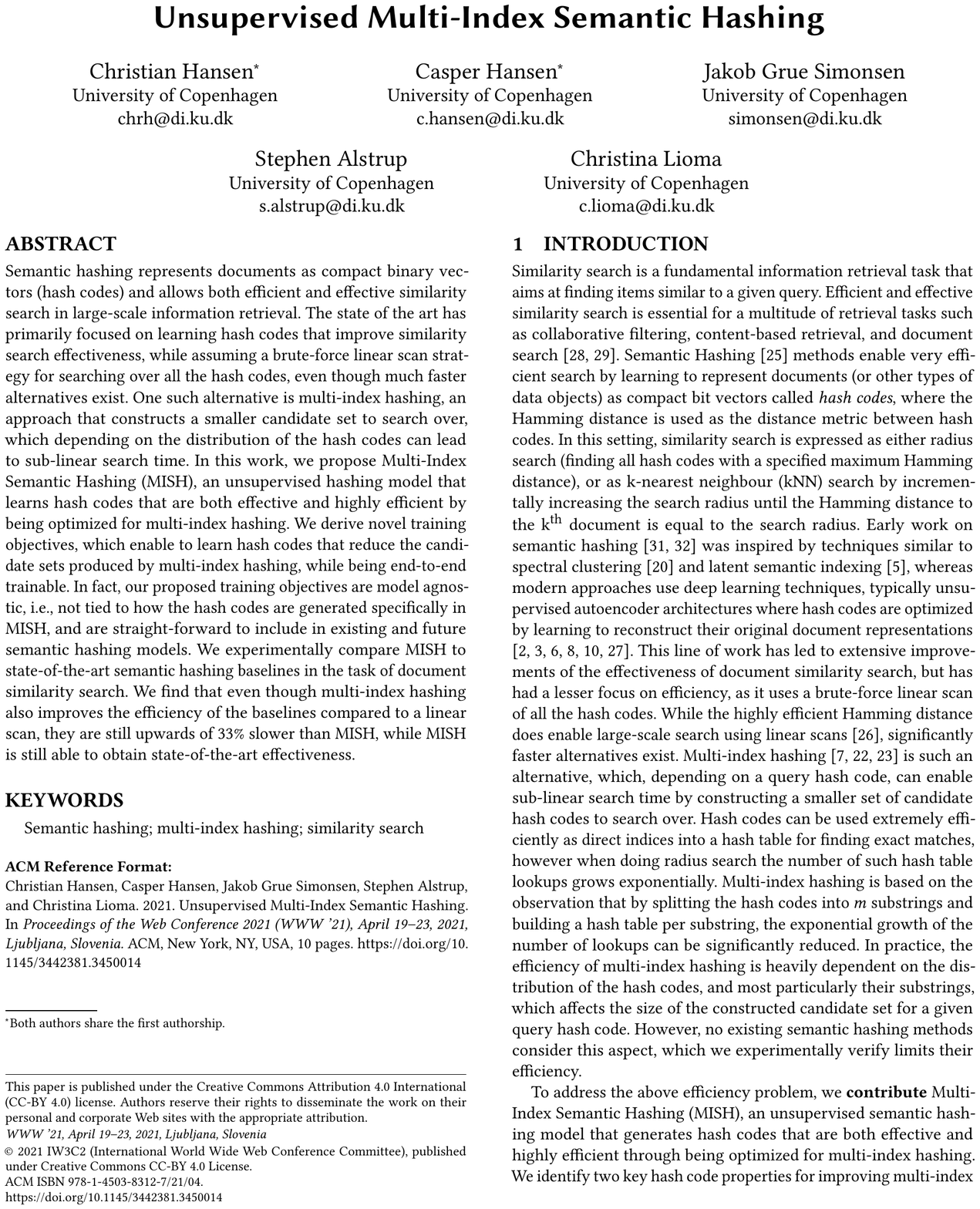}

\chapter{Content-aware Neural Hashing for Cold-start Recommendation} \label{chapter5}
Casper Hansen$^*$, Christian Hansen$^*$, Jakob Grue Simonsen, Stephen Alstrup, Christina Lioma (2020). Content-aware Neural Hashing for Cold-start Recommendation. In SIGIR, pages 971-980. \cite{DBLP:conf/sigir/hansen-sigir2020-coldstart}. $^*$ denotes equal contribution.
\includepdf[pages=-, pagecommand={}]{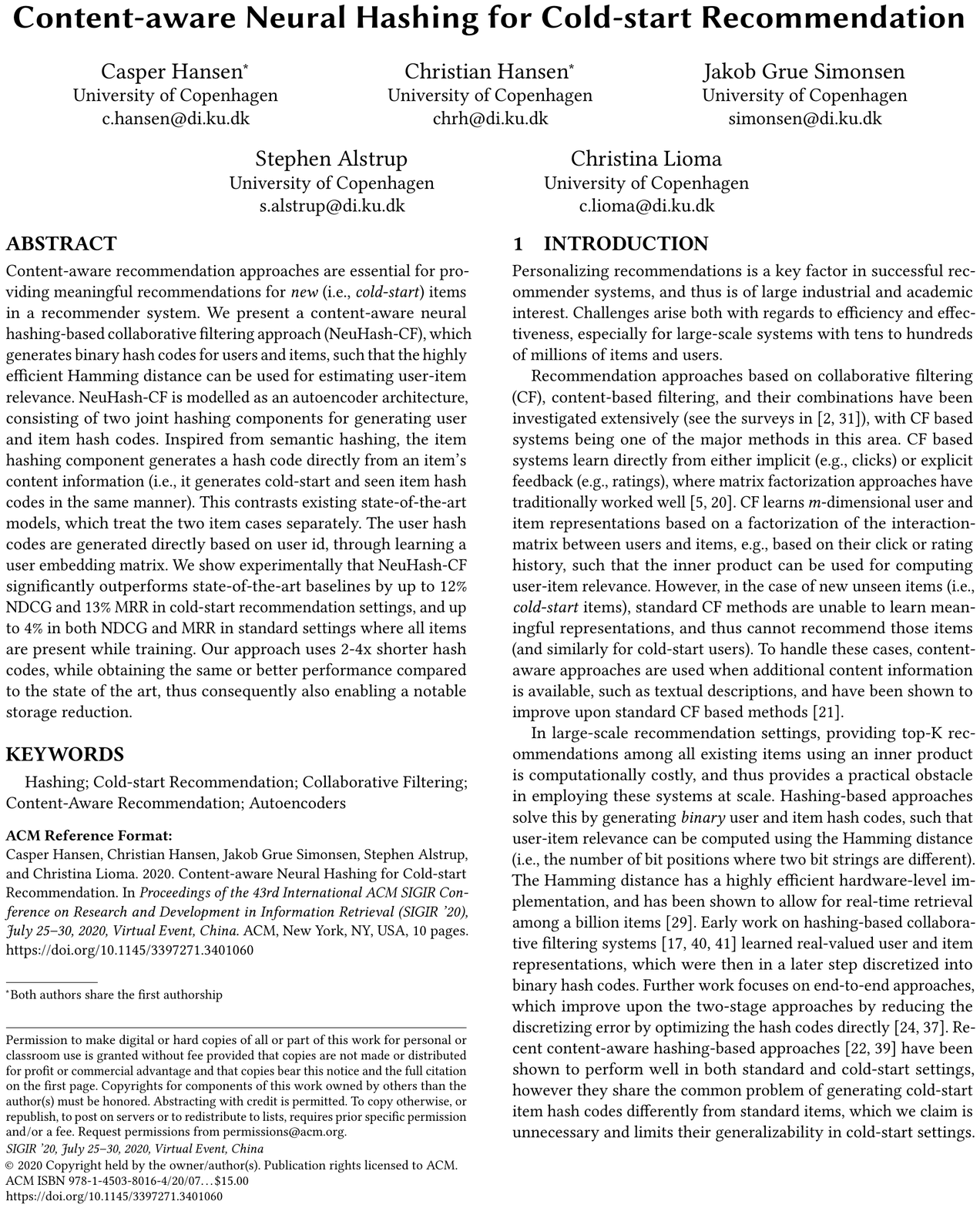}

\chapter{Projected Hamming Dissimilarity for Bit-Level Importance Coding in Collaborative Filtering}
\label{chapter6}
Christian Hansen$^*$, Casper Hansen$^*$, Jakob Grue Simonsen, Christina Lioma (2021). Projected Hamming Dissimilarity for Bit-Level Importance Coding in Collaborative Filtering. In WWW, in press. \cite{DBLP:conf/www/hansen-www2021-phd}. $^*$ denotes equal contribution.
\includepdf[pages=-, pagecommand={}]{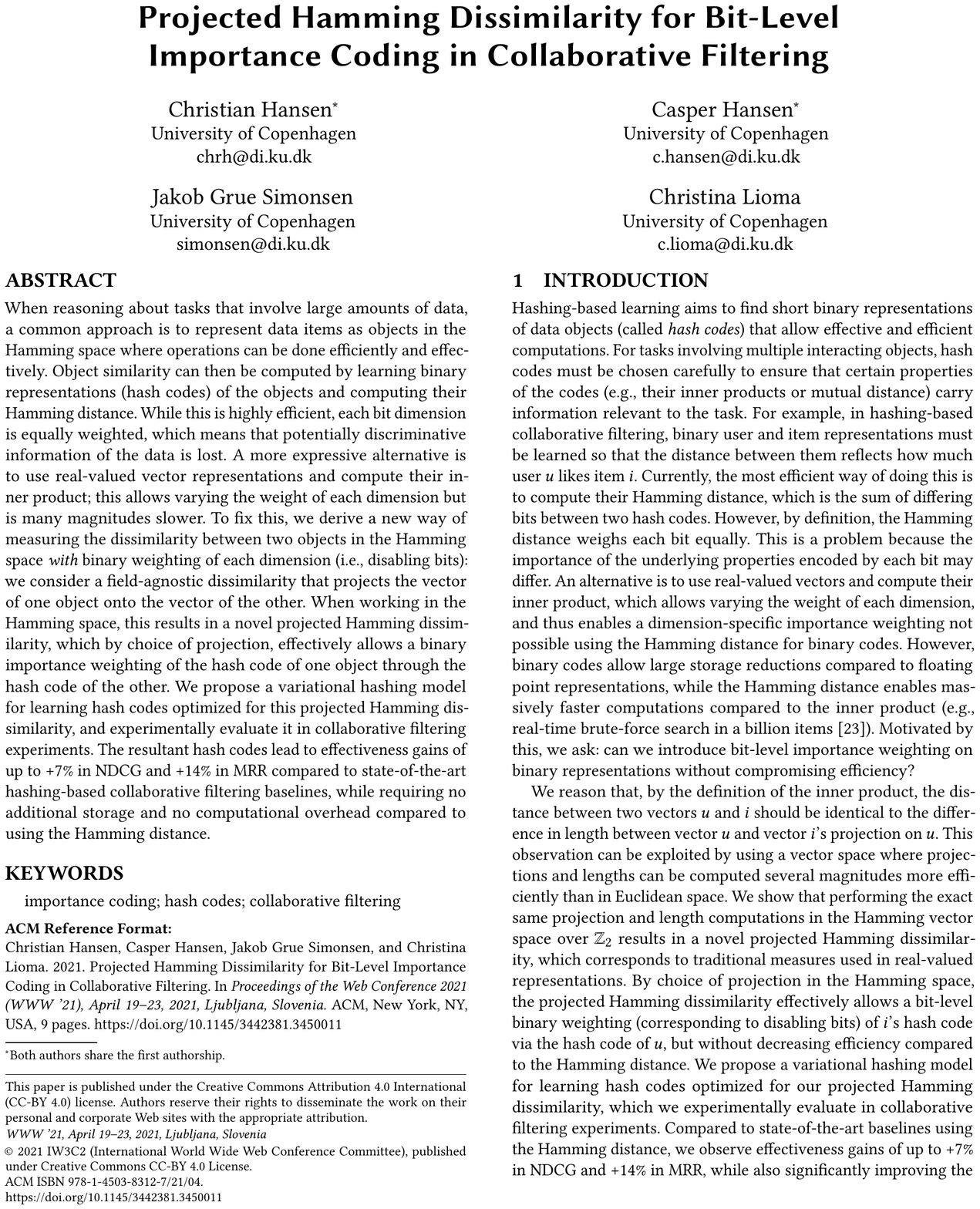}


\clearpage
\phantomsection
\addcontentsline{toc}{chapter}{Bibliography}
\bibliographystyle{plain}
\bibliography{references}

\end{document}